\documentclass[12pt]{article}
\usepackage{amssymb}

\input{tcilatex}
\begin{document}

\title{\textbf{WITTEN DEFORMED EXTERIOR DERIVATIVE AND BESSEL FUNCTIONS}}
\author{\textbf{MEKHFI.M } \\
{\normalsize Laboratoire de Physique Math\'{e}matique,Es-senia 31100 Oran 
\textbf{ALGERIE}}\\
{\normalsize and }\\
{\normalsize Int' Centre for theoretical Physics ,Trieste 34100 \textbf{ITALY%
}}}
\date{2000 April 18}
\maketitle

\begin{abstract}
In a recent paper we investigated the internal space of Bessel functions
associated with their orders.We found a formula (new) unifying Bessel
functions of integer and of real orders. In this paper we study the deformed
exterior derivative system $H=d_{\lambda }$ on the puctured plane as a
tentative to understand the origin of the formula and find that indeed
similar formula occurs.This is no coincidence as we will demonstrate that
generating functions of integer order reduced Bessel functions and of real
orders are respectively eigenstates of the usual exterior derivative and its
deformation .As a direct consequence we rediscover the unifying formula and
learn that the system linear in $d_{\lambda }$ is related to Bessel theory
much as the system quadratic in ($d_{\lambda }+d_{\lambda }^{*}$) is related
to Morse theory

{\footnotesize mekhfi@hotmail.com}
\end{abstract}

\begin{center}
\smallskip \newpage{}
\end{center}

\section{Introduction}

Early studies [e.g.,\cite{kn:Truesdell},\cite{kn:Infeld} ] proposed a
unifying scheme for special functions showing that some of these functions
may originate from the same structure .For Bessel functions of concern here
, generating functions of integer orders are representation ''states'' of
derivative and integral operators of arbitrary orders.More precisely we have
the ''inner'' structure

\begin{eqnarray}
\partial _{|m|} &=&\frac{\partial }{z\partial z}.\,\frac{\partial }{%
z\partial z}..........\frac{\partial }{z\partial z}.  \nonumber \\
\partial _{-|m|} &=&\int zdz.\,\int zdz..............\int zdz.  \nonumber \\
\partial _{m}\Phi (z,t) &=&(-\ t)^{-m}\,\,\Phi (z,t)\,\,,m\epsilon Z
\label{eq:1} \\
\Phi (z,t) &=&\sum_{n=-\infty }^{n=\infty }\phi _{n}(z)\ t^{n}  \nonumber
\end{eqnarray}

\noindent $\ $

\noindent where we extend the index m to negative values by introducing the
symbol $\int dz$ to denote a ``truncated '' primitive i.e. in defining the
integral we omit the constant of integration $\int \frac{df}{dz}dz=f$ and
where $\phi _{n}(z)$ stands for the reduced Bessel function $\phi _{n}(z)=%
\frac{J_{n}(z)}{z^{n}}$ of integer order. For the polynomials such as
Hermite and Laguerre for instance ,the generating functions only involve the
realization of the set N of positive integers,with slight modifications of
the derivative operators to account for the conventions used in defining
these polynomials.It is important to note that although this common
structure only set up the z-dependance of the generating functions ,it is
the `` dynamical '' part of the scheme so to say.The t dependence is simply
set by imposing some given desired properties.For Bessel functions for
example we require a ``symmetry'' between positive and negative integer
indices that is $J_{-n}=(-1)^{n}J_{n}$ ,while for the polynomials it is the
natural property of orthonormality that is invoked.

In a recent paper\cite{kn:Wissale} we intuitively applied a mechanism to
generate real numbers out of integers in order to unify ( reduced ) Bessel
functions and showed by direct analytic computation that indeed , Bessel
functions fit into the scheme and therefore integer orders are mapped to
real orders ( $\lambda $ is real ) through the formula 
\begin{equation}
\frac{J_{n+\lambda }(z)}{z^{n+\lambda }}=exp\left[ -\lambda \sum_{m\epsilon
Z/(0)}\,\frac{\,\partial _{m}}{m}\right] \frac{J_{n}(z)}{z^{n}}  \label{eq:2}
\end{equation}

\ 

Let us summarize the mechanism to see how it works to convert an integer
into a real.Suppose we are given an abstract state $\mid n\rangle
\,\,n\epsilon Z$ and a set of raising $\Pi _{m},m>0$ and lowering $m<0$
operators .Then it is easy to show , given that data , that the state $\mid
n+\lambda >$ is related to the state $\mid n>$ through the following formula

\begin{equation}
\mid n+\lambda >=exp\left[ -\lambda \sum_{m\epsilon Z/(0)}\frac{(-1)^{m}\Pi
_{m}}{m}\right] \mid n>  \label{eq:3}
\end{equation}
\noindent

Fourier transforming the $\mid n>$ state as 
\begin{equation}
\mid n>=\int_{-\pi }^{\pi }\frac{d\theta }{2\pi }e^{in\theta }\mid \theta >
\label{eq:4}
\end{equation}
\noindent

where the $\Pi _{m}$ operators act on the $\mid \theta >$ state by simple
multiplication by the factor $e^{im\theta }$ we get

\begin{eqnarray}
\mid n+\lambda > &=&\int_{-\pi }^{\pi }exp\left[ -\lambda \sum_{m\epsilon
Z/(0)}\frac{(-1)^{m}\Pi _{m}}{m}\right] \,\,e^{in\theta }\mid \theta >\frac{%
d\theta }{2\pi }  \nonumber \\
&=&\int_{-\pi }^{\pi }e^{i(n+\lambda )\theta }\mid \theta >\frac{d\theta }{%
2\pi }  \label{eq:5}
\end{eqnarray}
\noindent Compare states in \ref{eq:4} to states in \ref{eq:5}. In deriving
the last line use have been made of the known formula

\begin{equation}
\sum_{m=1}^{\infty }(-1)^{m}\frac{sinm\theta }{m}=-\frac{\theta }{2}%
,\,\,\,-\pi <\theta <\pi \,  \label{eq:6'}
\end{equation}

Formula \ref{eq:2} is a new formula ( to be added to the huge literature on
Bessel functions ) which is shown to apply to Neumann and Hankel functions
as well \cite{kn:Mohsine}.It is to be noted that although formula \ref{eq:2}
is shown to be true ,we didn't know why the above mechanism should apply to
Bessel functions.Our guess of the above relation was based on the following
correspondence \cite{kn:Wissale}

\begin{eqnarray*}
\phi _{n}(z) &\Leftrightarrow &\mid n> \\
\phi _{n+\lambda \text{ }}(z) &\Leftrightarrow &\mid n+\lambda >
\end{eqnarray*}

\ A hint to this correspondence came from the fact that we indeed have a set
of raising and lowering operators $\Pi _{m}=(-1)^{m}\partial
_{m}\,\,m\epsilon N$ \cite{kn:watson} \cite{kn:smirnov}

\begin{equation}
(-1)^{m}\frac{d^{m}}{(zdz)^{m}}\phi _{n}(z)=\phi _{n+m}(z)\hspace{5mm}%
m\epsilon N,n\epsilon Z  \label{eq:6}
\end{equation}
\noindent and for negative $m^{^{\prime }s}$ we just have to replace
derivative operators by integral operators defined in \ref{eq:1}.That this
correspondence works is quite intriguing and therefore further
investigations of it are needed .In this paper we answer the point. In
section 2 we will study the deformed exterior derivative on the punctured
plane and will realize that the converting( or deforming) mechanism is
closely tied to the deformation of the exterior derivative .In section 3 we
will demonstrate directly that real order reduced Bessel functions are the
deformed versions of integer order reduced Bessel functions in the same
fashion as $d_{\lambda }$ is the deformed version of the exterior derivative 
$d$ .As a consequence formula \ref{eq:2} will show up more elegantly.

\section{Deformed exterior derivative on $R^{2}/(0)$}

Let $M$ be a Riemannian manifold of dimension n .let $V_{p}$ $p=0,1....n$ be
the space of $p$-forms .Let $d$ and $d^{*\text{ }}$ be the usual exterior
derivative which define the De Rham cohomology of $M$ and its adjoint.Let $V 
$ be a smooth function $V:M\rightarrow R$ $or\ C$ ( called prepotential in
the language of topological quantum field theories ) and $\lambda $ a real
number .Define

\begin{equation}
d_{\lambda }[V]=e^{-\lambda V}d\,\,e^{\lambda V}  \label{eq:7}
\end{equation}
Evidently we have $d_{\lambda }^{2}=d_{\lambda }^{*2}=0$. E.Witten\cite
{kn:witten} had shown that $V$ plays the role of a Morse function and his
consideration of the system

\begin{equation}
H_{\lambda }=d_{\lambda }d_{\lambda }^{*}+d_{\lambda }^{*}d_{\lambda }
\label{eq:7'}
\end{equation}
had led to a new proof of Morse inequalities.Let us note at this point that
there exists another version of $d$-deformation which is related to the
fixed point theorems for Killing vector fields

\[
d_{s}=d+s\ iK 
\]
where s is an arbitrary number and where $K$ is a killing vector field-the
infinitesimal generator of an isometry of $M$.In this context $K$ is
regarded as an operator $iK$ on differential forms acting by interior
multiplication and hence maps a $p$-form into a $(p-1)$-form .Since we are
interested in functions ($0$-form) such a deformation is not relevant as $%
d_{s}$ coincide with $d$ on the space of functions.In this section and the
subsequent section we investigate the simpler system

\[
H=d_{\lambda } 
\]
and show that it gives informations on the index structure of Bessel
functions.Let us note at once that the above system is topological in the
sense that $d_{\lambda }$ ,like $d$ or $d_{s}$, can be defined purely in
terms of differential topology without choosing a metric in $M$ .Now to
proceed we need to know the appropriate form of $V.$The system in \ref{eq:7'}
has also been investigated , in another context , by Baulieu et all \cite
{kn:baulieu} to get informations on topological invariants.Their analysis of
topological quantum mechanics on the punctured plane $R^{2}/(0)$ had
selected the prepotential $V=k\theta $ which we later generalized .We have
shown \cite{kn:Nadia}that the most general prepotential compatible with the
topology of the punctured plane ( first homotopy group $\sim $ $Z$ ) has
necessary the form.

\begin{equation}
V(\theta )=k\theta +\phi (\theta )  \label{eq:8}
\end{equation}
\noindent where $\theta $ is the polar angle on the plane , k a constant and 
$\phi (\theta )$ any function but periodic , (recall that the polar angle $%
\theta $ is not a periodic function ).It is that form \ref{eq:8} that we
plug into $d_{\lambda \text{ }}$.On the restricted space of functions which
depend only on the angle ,the exterior derivative simplifies to $d=d\theta
\partial _{\theta }\ $( there is no r dependance on which d acts )
.Inserting the specific form of the prepotential $V$ into ~\ref{eq:7} and
rewriting the twisted operator as $d_{\lambda }=d\theta \partial _{\theta
}^{\lambda }$ we find

\[
\partial _{\theta }^{\lambda }=e^{-\lambda \phi }\partial _{\theta
}\,\,e^{\lambda \phi }+\lambda k=\partial _{\theta }+\lambda k+\lambda
\partial _{\theta }\phi 
\]
\noindent Fourier transforming the periodic function $\phi (\theta )$ 
\[
\partial _{\theta }\phi (\theta )=i\sum_{m\epsilon Z/0}\rho _{m}\
e^{-im\theta } 
\]
\noindent we get 
\begin{equation}
\partial _{\theta }^{\lambda }=\partial _{\theta }+i\lambda \sum_{m\epsilon
Z}\rho _{m}e^{-im\theta }\noindent  \label{eq:9}
\end{equation}
where we inserted the constant $k=i\rho _{0}$ into the sum .In the punctured
plane the operator $\partial _{\theta }$ and $\partial _{\theta }^{\lambda }$
have the natural interpretation respectively of the winding number operator
and the effective or perturbed winding number operator .We thus\quad write
them as $W=-i\partial _{\theta }$ and $\,W_{\lambda }=-i\partial _{\theta
}^{\lambda }$.$\,$We also introduce the operator $\Pi _{m}=e^{im\theta }$
with evident action on the basis $\mid n>$ defined in \ref{eq:4} . For the
operator $W$ and $\Pi _{m}$ we have $W\mid n>=n\mid n>$ and $\Pi _{m}\mid
n>=\mid n+m>$.In this new basis \ref{eq:9} takes the form

\[
W_{\lambda }=W+\lambda \sum_{m\epsilon Z}\rho _{m}\Pi _{m} 
\]

This is an example of a very simple topological quantum mechanical system
where $W_{\lambda }$ is the perturbed hamiltonian ,$\Pi _{m}$ a set of
operators responsible for the interactions, $\lambda \rho _{m}$ a set of
coupling constants and $W$ is the unperturbed hamiltonian .The eigenstates
of the effective winding number which we denote $\mid n,\lambda $ $,\rho >$
are shown to be related to the unperturbed one ,through the formula \cite
{kn:Najoua} 
\begin{equation}
\mid n,\lambda \text{ },\rho >=exp\left[ -\lambda \sum_{m\epsilon Z/(0)}%
\frac{\rho _{m}\Pi _{m}}{m}\right] \mid n>  \label{eq:10}
\end{equation}
\noindent Hermiticity of $W_{\lambda }$ restricts the real `` spectral ''
function $\rho $ to be symmetric $\rho _{m}=\rho _{-m}$ .Comparing with the
previous result \ref{eq:3} we see that our application of the formalism to
Bessel functions requires the simple choice of the function $\rho
_{m}=(-1)^{m}$ .

\section{Relation of $\phi _{n+\lambda }\ $to $d_{\lambda }$}

To show the relation , first write the generating functions of integer and
of real orders

\begin{eqnarray}
\Phi (z,t) &=&\sum_{n=-\infty }^{n=\infty }\phi _{n}(z)\ t^{n}  \nonumber \\
t^{-\lambda }\ \Phi (z,t) &=&\sum_{n=-\infty }^{n=\infty }\phi _{n+\lambda
}(z)\ t^{n}  \label{eq:11}
\end{eqnarray}
We have learned in the particular case of section 2 that eigenstates of $%
d_{\lambda }$ ( $W_{\lambda }\ $) are deformed versions of the eigenstates
of $d$ ( $W$ )and that the deformation consists in converting the index $n$
into $n+\lambda $ .We therefore have to look for eigenstates of $d$ and of $%
d_{\lambda }$ .The generating function $\Phi (z,t)$ $\ $is an eigenstate of
the exterior derivative by use of the recursion formula ( fixed t )
\[
d\ \Phi (z,t)=(-\frac{t}{z})^{-1}\ \Phi (z,t)\ dz
\]
For the eigenstate of $d_{\lambda }$ , the function of interest to look at
is $e^{-\lambda V}\Phi (z,t)$ .We will show that ,with an appropriate $V$
,it is indeed an eigenstate of the deformed exterior derivative and in the
same time generating function of real order Bessel functions.The judicious
choice of the operator $V$ so as to identify this function with the
generating function for real orders (remember that the same crucial point of
which $V$ to choose arose in the last section) can be shown to be \-%
\footnote{%
In choosing a ( differential ) operator for $V$ instead of a simple function
as in \ref{eq:7} we have implicitly generalized the deformed exterior
derivative on flat space.This is enough for our purpose .We do not however,
know , the expression of the generalized $d_{\lambda }$ in the case of a
general manifod $M$.Such expression should be defined so as to be covariant
and not to depend on the metric on $M$ like $d=dzD_{z}=dz\partial _{z}.$}

\[
V=\sum_{m\epsilon Z/0}\frac{\partial _{m}}{m} 
\]

In fact we have

\begin{eqnarray}
e^{-\lambda V}\Phi &=&\exp \left[ -\lambda \sum_{m\epsilon Z/0}\frac{%
\partial _{m}}{m}\ \right] \Phi  \nonumber \\
&=&\exp \left[ -\lambda \sum_{m\epsilon Z/0}\frac{(-1)^{m}(t)^{-m}}{m}%
\right] \Phi  \label{eq:12} \\
&=&t^{-\lambda \ }\Phi (z,t)  \nonumber
\end{eqnarray}

\smallskip This is the generating function of real order Bessel functions.To
come to the second line we applied the recursion formula \ref{eq:1} and from
the second line to the last we put $t=e^{i\theta \text{ }}$and made use of 
\ref{eq:6'}.Then acting by $d_{\lambda }$ on $e^{-\lambda V}\Phi (z,t)$ we
find

\begin{equation}
d_{\lambda }(e^{-\lambda V}\Phi )=e^{-\lambda V}d\Phi =(-t)^{-1}dz\
e^{-\lambda V}(z\ \Phi (z,t))=\digamma (z,t,\lambda )\ dz\ e^{-\lambda V}\Phi
\label{eq:13}
\end{equation}

The function $\digamma (z,t,\lambda )$ is a more involved expression which
we will not work out as this is not necessary .Inspection of the third term
in \ref{eq:13} shows that $e^{-\lambda V}(z\ \Phi )\backsim \ \Phi \backsim
e^{-\lambda V}\Phi \ $where the last proportionality comes from the result
in \ref{eq:12} .

Thus using the fact that the generating function of integer orders $\Phi
(z,t)$ is an eigenstate of $d$ ,we show that the function $e^{-\lambda
V}\Phi $ is indeed an eigenstate of $d_{\lambda }$ and generating function
of real orders .

Expanding both sides of \ref{eq:12} as in \ref{eq:11} the above relationship
extends to Bessel functions themselves as the operator $V$ acts only on the $%
z$ variable.Hence we recover the unifying formula

\[
\phi _{n+\lambda }(z)=\exp \left[ -\lambda \sum_{m\epsilon Z/0}\frac{%
\partial _{m}}{m}\right] \phi _{n}(z) 
\]

The method outlined in this section has the advantage of giving a new check
to the unifying formula ,in addition it shades light on the inner structure
of Bessel functions showing that the modern concept of deformed or twisted
exterior derivative ,first introduced by Witten ( which has been at the
origin of the launching of topological field theories ) is already encoded
in the index structure of Bessel functions .

\begin{quotation}
\textbf{Acknowledgment}
\end{quotation}

I would like to personally thank the head of the high energy section at the 
\textbf{Abdus Salam }international centre for theoretical physics ICTP Dr
S.Randjbar -Daemi for inviting me to the centre scientific activities at
various occasions.\-\newpage {}

\end{document}